\documentclass[%
 reprint,
superscriptaddress,
amsmath,amssymb,
aps,
pra,
]{revtex4-1}
\usepackage{caption}
\usepackage{subcaption}
\usepackage{graphicx}
\usepackage{xcolor}
\usepackage{graphicx}
\usepackage{dcolumn}
\usepackage{bm}
\usepackage{upgreek}
\usepackage{lineno}

\begin{document}

\preprint{APS/123-QED}
\title{Compact dual-band spectral analysis via\\  multiplexed rotated chirped volume Bragg gratings}

\author{Oussama Mhibik}
\affiliation{CREOL, The College of Optics \& Photonics, University of Central Florida, Orlando, FL 32816, USA}
\author{Murat Yessenov}
\thanks{Corresponding author: yessenov@knights.ucf.edu}
\affiliation{CREOL, The College of Optics \& Photonics, University of Central Florida, Orlando, FL 32816, USA}
\author{Leonid Glebov}
\affiliation{CREOL, The College of Optics \& Photonics, University of Central Florida, Orlando, FL 32816, USA}
\author{Ayman F. Abouraddy}
\affiliation{CREOL, The College of Optics \& Photonics, University of Central Florida, Orlando, FL 32816, USA}
\author{Ivan Divliansky }
\affiliation{CREOL, The College of Optics \& Photonics, University of Central Florida, Orlando, FL 32816, USA}

\begin{abstract}
Chirped Bragg volume gratings (CBGs) offer a useful alternative for spectral analysis, but increasing the bandwidth necessitates increasing the device area. In contrast, recently developed rotated CBGs (r-CBGs), in which the Bragg structure is rotated by $45^{\circ}$ with respect to the device facets, require increasing only the device length to extend the bandwidth, in addition to the convenience of resolving the spectrum at normal incidence. Here, we multiplex r-CBGs in the same device to enable spectral analysis in two independent spectral windows without increasing the system volume. This new device, which we term an X-CBG, allows for compact multi-band spectroscopy in contiguous or separated spectral windows in the visible and near-infrared for applications in nonlinear microscopy and materials identification and sensing.
\end{abstract}



\maketitle

Spectrometry is one of the fundamental tools used in optics for the identification of materials and chemical compounds, for monitoring changes in the environment, for quality control of food and industrial processes, and for medical diagnostics \cite{Parson2007Book}. Optical spectral analysis can be implemented via diverse strategies, including tunable narrowband filters \cite{Gat2000Review}, Fourier transform systems \cite{Saptari2003Book}, and -- more recently -- by exploiting computational techniques for reconstructing the spectrum by trained detectors \cite{Redding13NP}. The leading approach for many commercial systems in the visible and near-infrared (NIR) remains the utilization of a dispersive element, such as diffraction gratings, to spatially resolve the spectrum before recording it with a linear detector array \cite{Tkachenko2006Book}. Such systems occupy a middle ground with respect to system performance, cost, and size.

Over the past three decades, efforts have been dedicated towards miniaturization of the volume of spectrometers \cite{Yang21Science}, which enables applications that require handheld or portable devices \cite{Crocombe18AS,Bacon04RSI} for use in smartphones \cite{McGonigle18Sensors}, for hyperspectral imagers on drones, or for on-chip implementations \cite{Cheng19NC,Yang19Science} (see the recent reviews in Refs.~\cite{Yang21Science,Li22LSA}). One example exploits random media in lieu of a conventional diffraction grating, whether a multimode fiber \cite{Redding13OE} or an on-chip structure \cite{Redding13NP,Hartman20AOM}. Another recent example combines computational techniques with new measurement techniques to drastically reduce the system volume \cite{Yoon22Science}. It is nevertheless now well-understood that any optical functionality, such as spectral analysis, requires a minimal volume to be carried out \cite{Miller07JOSABFundamental,Miller22Thickness}. In other words, even if the optical components used are reduced to thin elements (e.g., metasurfaces), a minimum system volume is nevertheless required. This volume is usually dominated by the space needed for free propagation rather than the optical components themselves.

We recently introduced a new class of compact optical devices for spectral analysis that we have called rotated chirped volume Bragg gratings (r-CBGs) \cite{Mhibik23OL}. In an r-CBG, the Bragg structure is rotated by $45^{\circ}$ with respect to the device facets. Consequently, the spectrum of a field incident \textit{normally} on the input facet is spatially resolved and exits normally from a side facet orthogonal to the input. Such a device enables constructing compact, mechanically stable spectrometers. Moreover, r-CBGs have been shown to facilitate ultra-compact systems for spatio-temporally structuring light, which has been validated by synthesizing space-time wave packets \cite{Yessenov22AOP,Yessenov23OL}.

Despite the rich tapestry of available spectroscopic techniques, a useful degree-of-freedom has yet to be realized to the best of our knowledge; namely, multiplexing multiple spectrometers \textit{in the same volume}, each operating in a different spectral range and with potentially different bandwidths. Of course, two different spectrometers could be combined, but this adds to the system volume and complexity. Providing spectral analysis simultaneously in two different ranges in the same compact device with no need for first separating the input signals has important applications in pump-probe measurements, and fluorescence and nonlinear microscopy. In many of these scenarios, two optical signals in different spectral ranges but sharing a common path are of interest.

In this paper, we report on a new advance in the functionality of CBGs by realizing a dual-band spectral analysis device in which two r-CBGs are multiplexed in the same volume. One r-CBG is written at an angle $45^{\circ}$ with respect to the propagation axis and operates in one spectral range, and another r-CBG operating in a distinct spectral range is written at an angle $-45^{\circ}$ (orthogonal to the first). Because the chirped Bragg structure has the form of orthogonally intersecting lines throughout the device volume, we call this new optical element an X-CBG. When two optical signals in different spectral bands are incident normally at the entrance facet, the spectrum of one signal is spatially resolved and directed normally out of one facet, while the spatially resolved spectrum of the other signal exits from a different facet. We confirm the operation of X-CBG's in visible and NIR spectral channels, and in contiguous spectral ranges in the visible. We verify that the parameters of each spectral channel (central wavelength and bandwidth) can be tuned independently by implementing different chirp profiles. This work paves the way to compact multi-channel spectral analysis in the visible and the near-infrared (NIR).

\begin{figure}[t!]
    \centering
    \includegraphics[width=8.6cm]{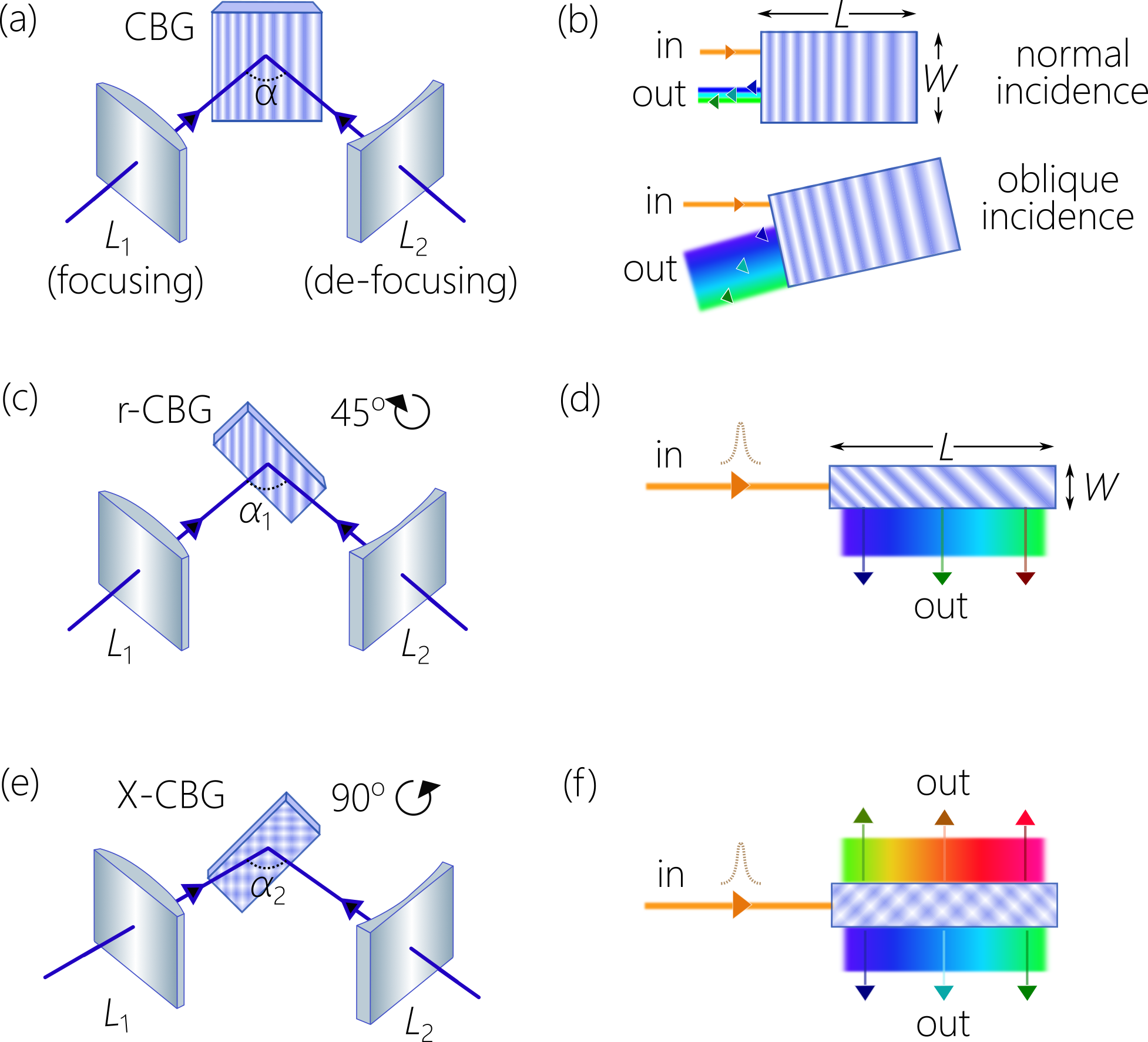}
    \caption{(a) Schematic of the writing procedure for a conventional CBG. Two overlapping laser beams produced by focusing and de-focusing lenses result in an interference pattern. (b) Mode of operation of a CBG: a normally incident pulse on a CBG of dimensions $L\times W$ is temporally stretched, and spectral analysis is achieved only for an obliquely incident signal. (c,d) Same as (a) for an r-CBG. Spectral analysis can be achieved here at normal incidence. The width $W$ can be reduced dramatically with respect to the corresponding dimension of the conventional CBG in (a,b). (e) By multiplexing a second r-CBG in the same sample from (c) after a $90^{\circ}$-rotation and change in the relative angle between the two writing beams to shift the spectral band, we obtain an X-CBG. The dimensions of the X-CBG are the same as those of the r-CBG in (c,d). (f) Each r-CBG multiplexed in the X-CBG produces an independent spectral channel in which the spectrum is spatially resolved.}
    \label{fig:Concept}
\end{figure}

We start by describing the writing process and the structure of X-CBGs in comparison to traditional CBGs and the recently introduced r-CBGs, which is illustrated in Fig.~\ref{fig:Concept}. A pair of UV laser beams (one converging and the other diverging) are combined at a prescribed relative angle to produce an interference pattern and write the target chirped Bragg structure in a photosensitive sample \cite{Glebov2009PTR, Nikolay17PTR}. This results in a conventional CBG [Fig.~\ref{fig:Concept}(a)] in which the Bragg structure is parallel to the sample facets \cite{Ciapurin12OEng}. The central wavelength is changed by tuning the relative angle between the two overlapping beams, and the chirp rate is changed by varying the focal lengths of the two lenses producing the converging and diverging beams \cite{Benoit2018Chirp}.

A normally incident field is retro-reflected after acquiring a spectral chirp and is thus stretched temporally [Fig.~\ref{fig:Concept}(b)] because each wavelength is reflected from a different depth within the structure \cite{Kaim14OEng,Glebov14OEng}. This configuration is useful for coherent pulse amplification systems \cite{Glebov14OEng,Liao07OE}. Such a CBG can be operated in a different modality to spatially resolve the spectrum of the incident field. A collimated, obliquely incident field is reflected with the spatially resolved spectrum \cite{Yessenov22NC} -- so-called spatial dispersion \cite{Gerken03IEEEPTL}, which is useful in optical communications for signal multiplexing and de-multiplexing; see Fig.~\ref{fig:Concept}(b). As mentioned earlier, increasing the bandwidth at a fixed chirp rate requires increasing both the length $L$ and the width $W$ \cite{Mhibik23OL}. It is easy to see that increasing the resolved bandwidth at a fixed chirp rate requires increasing the \textit{area} of the CBG \cite{Yessenov23CompactSTWP}. 

An r-CBG can be written using the same interference pattern after rotating it by $45^{\circ}$ with respect to the sample volume [Fig.~\ref{fig:Concept}(c)]. For a normally incident field, different wavelengths reflect from different positions along the device and exit normally from a facet orthogonal to the input [Fig.~\ref{fig:Concept}(d)]. By virtue of this novel geometric configuration, increasing the bandwidth of an r-CBG at fixed chirp rate necessitates increasing only the \textit{length} of the device $L$ rather than its area. This dramatically reduces the volume required for resolving a target bandwidth with respect to a conventional CBG \cite{Mhibik23OL}. Moreover, the spectrum is resolved immediately at the r-CBG exit without need for any further propagation. This feature, in addition to the convenience of normal incidence and exit, allows for constructing ultra-compact, mechanically stable spectrometers by abutting a one-dimensional calibrated detector array to the r-CBG exit facet. 

An X-CBG is produced by first writing an r-CBG in the sample [Fig.~\ref{fig:Concept}(c)] and then rotating the sample by $90^{\circ}$ followed by recording a second r-CBG using a new interference pattern [Fig.~\ref{fig:Concept}(e)]. The two interferograms can be designed independently to produce chirped Bragg structures in different spectral bands and different bandwidths. Therefore, two independent but spatially overlapping Bragg structures are recorded in the same volume orthogonally to each other: one r-CBG is rotated by $45^{\circ}$ with respect to the sample axis length $L$, and the other r-CBG is rotated by $-45^{\circ}$. 

When a field is incident normally on the input facet, part of the spectrum is spatially resolved by one r-CBG and is directed to exit one facet, while a different spectral range is spatially resolved by the second r-CBG and exits the opposing facet [Fig.~\ref{fig:Concept}(f)]. Because the two Bragg structures are written independently of each other, the two resolved spectral bands can be -- in principle -- chosen arbitrarily (within the transparency window of the sample material used). For example, one r-CBG can be designed to resolve the visible spectrum and the other to resolve the NIR. Alternatively, the two r-CBGs can be designed to resolve two contiguous spectral ranges. Both of these scenarios will be realized experimentally below.

\begin{figure}[t!]
    \centering
    \includegraphics[width=8.6cm]{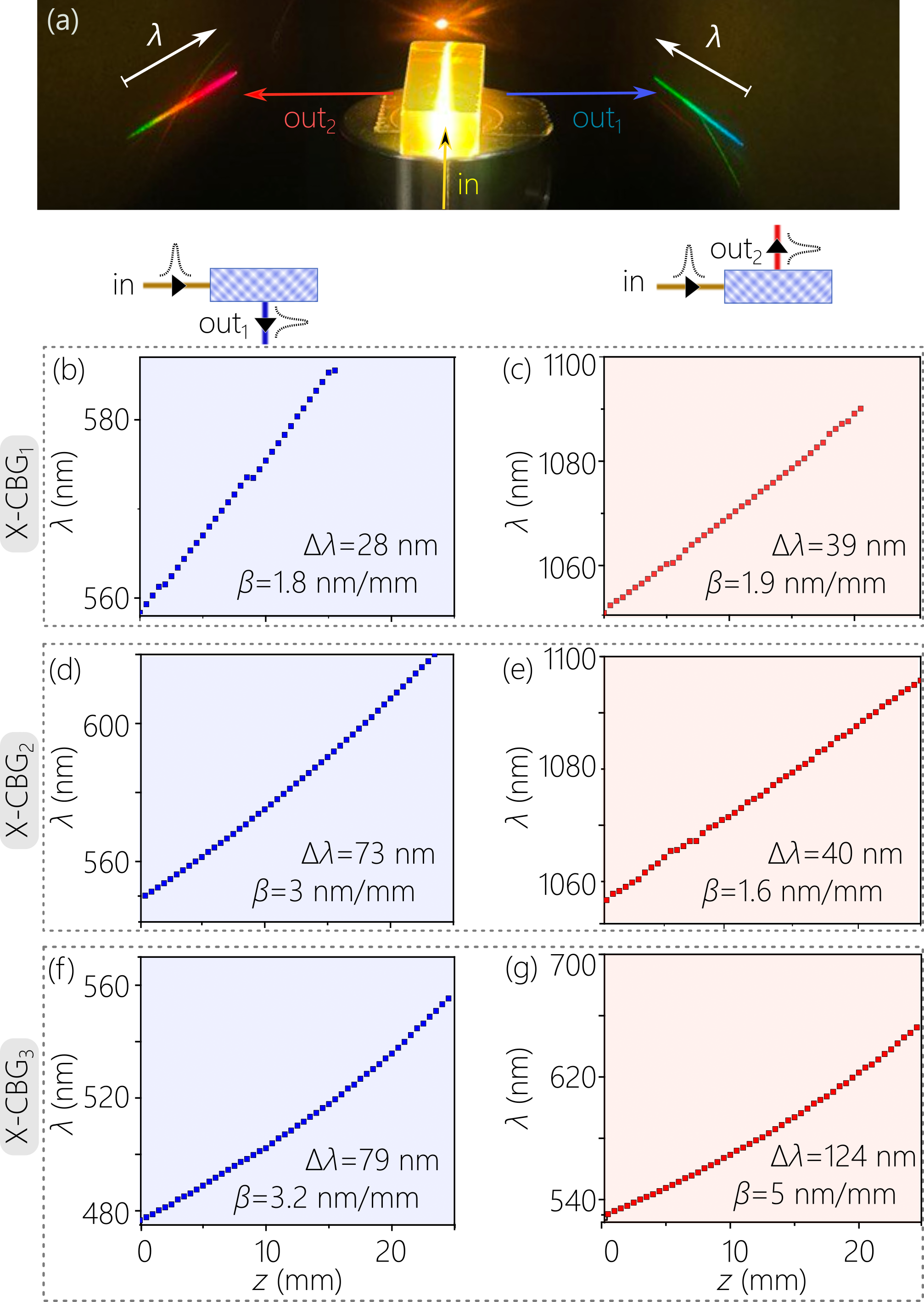}
    \caption{(a) Photograph of an X-CBG capable of dual-band spectral analysis in the visible. A white-light source is directed to the X-CBG (bottom), the spatially resolved short-wavelength portion of the spectrum is directed to the right, and the long-wavelength portion to the left. (b,c) Spectral measurement of the spatially resolved spectra from the two exit facets of X-CBG$_{1}$. (d,e) Same as (b-c) for X-CBG$_{2}$. (d) Same as (f,g) for X-CBG$_{3}$. The insets depict the measurement configurations.}
    \label{fig:SpectralMeasurement}
\end{figure}

We recorded the gratings in Photo-Thermo-Refractive glass (PTR) \cite{Nikolay17PTR} using a He:Cd UV laser at a wavelength 325~nm via a pair of positive and negative cylindrical lenses. The focal lengths are $f\!=\!\pm100$~cm for both outputs in X-CBG$_{1}$ [Fig.~\ref{fig:SpectralMeasurement} (b,c)]; $f\!=\!\pm50$~cm for output-1 of X-CBG$_{2}$ [Fig.~\ref{fig:SpectralMeasurement} (d)] and $f\!=\!\pm100$~cm for output-2 [Fig.~\ref{fig:SpectralMeasurement} (e)]; $f\!=\!\pm25$~cm for both outputs in X-CBG$_{3}$ [Fig.~\ref{fig:SpectralMeasurement} (f,g)].
For each r-CBG, the angle between the two focused beams was varied to adjust for the target wavelength. 

To characterize the X-CBGs, we used two broadband sources. A supercontinuum source (SuperK EVO ERL-04, NKT  Inc.) was used to characterize the visible grating, while a superluminescent diode (SLD1050S-A60) was used to characterize the NIR grating. Capturing the light field with a multimode fiber of 100-$\mu$m core diameter, we record the spectra of the visible and NIR spectra using two different spectrometers (Ocean Optics, S2000 and Thorlabs, CCS175, respectively). Scanning the fiber along the spatially resolved fields emerging from the two X-CBG spectral channels, we determined the chirp rate and the bandwidth of the two multiplexed r-CBGs. We produced here three X-CBGs whose spectra for both channels are plotted in Fig.~\ref{fig:SpectralMeasurement}(b-g). The lengths of all three devices are $L\!\approx\!25$~mm, and the input facet dimensions are $12\times 6$~mm$^{2}$.

In the first device, X-CBG$_{1}$, we introduced an r-CBG operating in the visible (bandwidth $\Delta\lambda\!\approx\!28$~nm) and another operating in the NIR ($\Delta\lambda\!\approx\!39$~nm). The measured chirp rates for the visible and NIR spectral channels are $\beta\!\approx\!1.8$~nm/mm [Fig.~\ref{fig:SpectralMeasurement}(b)] and $\beta\!\approx\!1.9$~nm/mm [Fig.~\ref{fig:SpectralMeasurement}(c)], respectively. In a second device, X-CBG$_{2}$, we maintained the central wavelengths of the visible and NIR spectral channels as in X-CBG$_{1}$ but increased the bandwidth of the visible channel ($\Delta\lambda\!\approx\!73$~nm) by increasing the chirp rate to $\beta\!\approx\!3$~nm/mm [Fig.~\ref{fig:SpectralMeasurement}(d,e)]. In the third device, X-CBG$_{3}$, we maintained the bandwidth of the first spectral channel but shifted the central wavelength of the NIR channel into the visible and increased the bandwidth ($\Delta\lambda\!\approx\!124$~nm, $\beta\!\approx\!5$~nm/mm). The two channels thus occupy contiguous portions of the visible spectrum [Fig.~\ref{fig:SpectralMeasurement}(f,g)].

To evaluate the spectral resolution of the X-CBGs, we illuminate the device with a broadband source and then scan a fiber along the spatially resolved spectrum. We carried out measurements in both spectral channels of X-CBG$_{2}$ [Fig.~\ref{fig:SpectralMeasurement}(d,e)]. The spectra were recorded with commercial spectrometers: Thorlabs CCS175 with a 1-nm resolution for the visible channel, and an optical spectrum analyzer (Yokogawa AQ6370D) with spectral resolution $\!\approx\!20$~pm in the NIR channel. We fix the fiber location and reduce the fiber core size, which results in a narrowing of the recorded spectrum. We find the spectrum width to reach a minimum value at a fiber core size of 100~$\mu$m: 1~nm for the visible channel and 0.5~nm for the NIR channel. These measurements reveal the intrinsic spectral resolution of the X-CBG in the two channels. Next, we scan the fiber to determine the spectral chirp rate along the $z$-axis (along which the spectra are spatially resolved). The FWHM of the spectral linewidth is 0.5~mm and 0.3~mm in the visible and NIR channels.

\begin{figure}[t!]
    \centering
    \includegraphics[width=8.6cm]{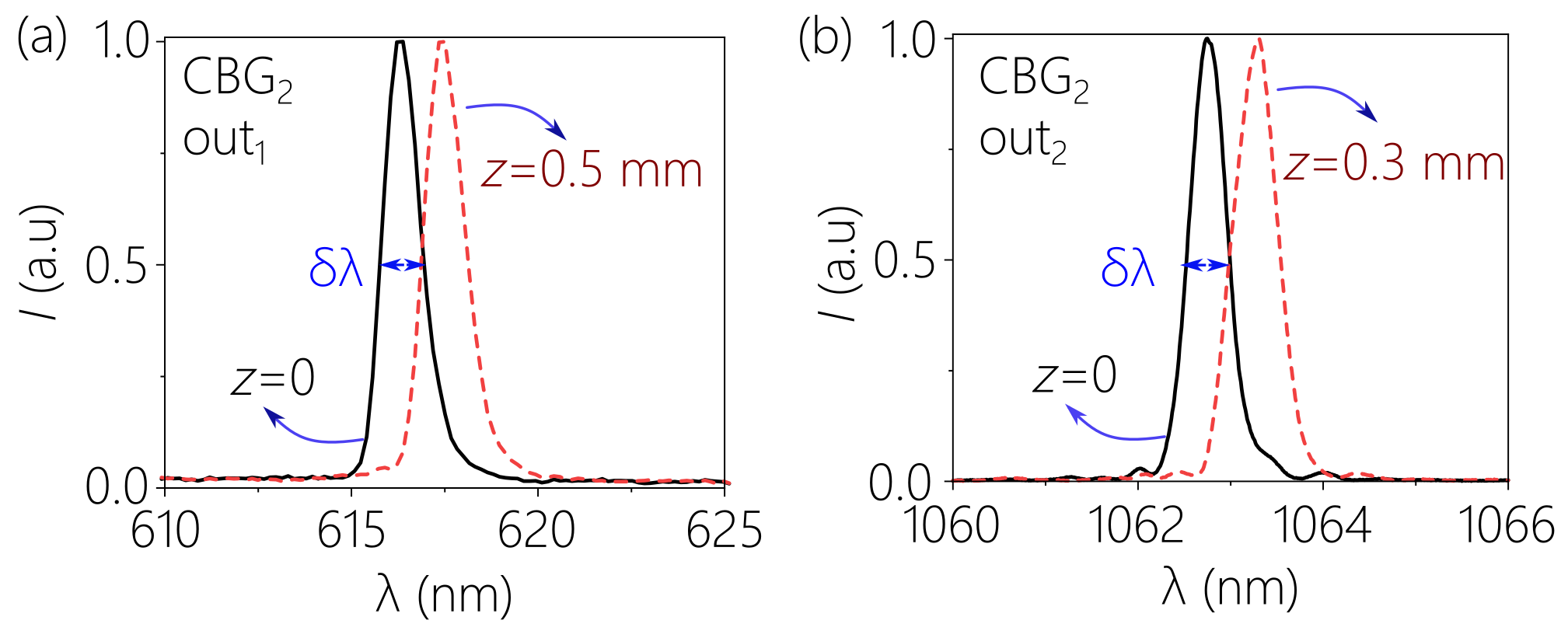}
    \caption{Evaluation of the spectral resolution of an X-CBG, corresponding to X-CBG$_{2}$ in Fig.~\ref{fig:SpectralMeasurement}(d,e). (a) Spectral resolution in the visible channel [Fig.~\ref{fig:SpectralMeasurement}(d)], and (b) spectral resolution in the NIR channel [Fig.~\ref{fig:SpectralMeasurement}(e)]. The solid and dotted spectra are captured by a multimode fiber at two neighboring locations along the $z$-axis.}. 
    \label{fig:SpectralResolution}
\end{figure}

\begin{figure}[t!]
    \centering
    \includegraphics[width=8.6cm]{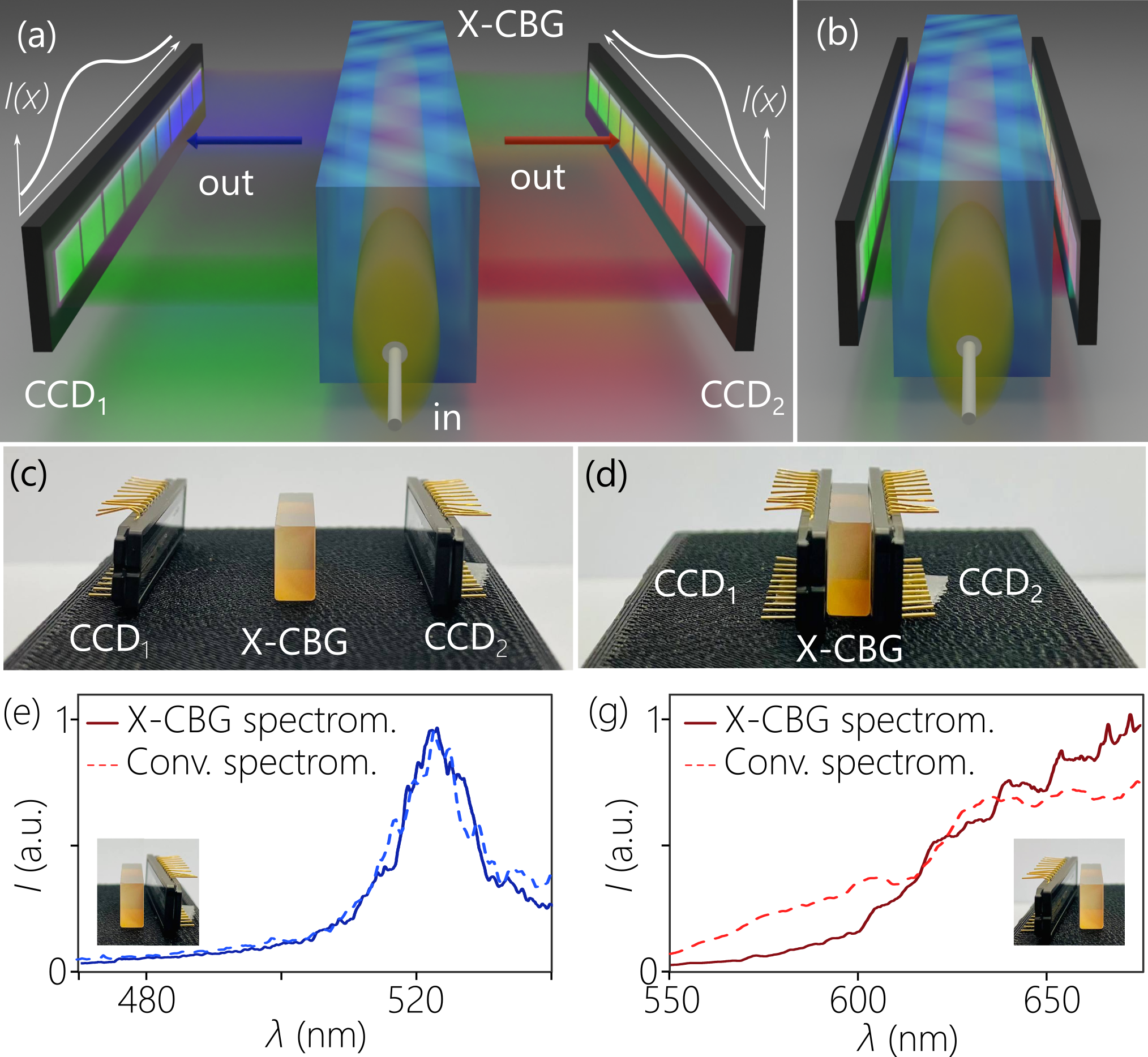}
    \caption{(a) Schematic of an envisioned compact dual-band spectrometer comprising an X-CBG and two CCD chips to intercept the spatially resolved spectra emerging to the left and right. (b) The two CCD chips can be abutted directly to the two output facets of the X-CBG. (c,d) Photographs of X-CBG$_{3}$ [Fig.~\ref{fig:SpectralMeasurement}(f,g)] with two linear silicon CCD arrays corresponding to the schematic diagrams in (a,b) for spectral dual-band analysis in the visible. (e,d) Measured spectra as captured by the linear CCD arrays in the two spectral bands of X-CBG$_{3}$ compared to a reference measurement captured with a commercially available spectrometer (Ocean Optics, S2000).}
    \label{fig:SystemPhotograph}
\end{figure}

The X-CBG is expected to lead to miniaturized spectroscopic devices. In Fig.~\ref{fig:SystemPhotograph}(a,b) we depict schematically the envisioned compact dual-band spectrometer enabled by an X-CBG. The field is incident normally at the X-CBG input facet, and two prescribed spectral ranges are directed to the left and to the right normally to the exit facets. Two wavelength-appropriate linear detector arrays are placed at these facets to intercept the spatially resolved spectra [Fig.~\ref{fig:SystemPhotograph}(a)]. Because the spectra are spatially resolved with no additionally required free-space propagation, the two arrays can be directly abutted to the X-CBG, resulting in an ultra-compact dual-band spectrometer system [Fig.~\ref{fig:SystemPhotograph}(b)]. 

In Fig.~\ref{fig:SystemPhotograph}(c,d) we demonstrate one possibility using the X-CBG from Fig.~\ref{fig:SpectralMeasurement}(f,g) that resolves two portions of the visible spectrum. We make use of two linear silicon CCD chips each comprising 3648~pixels (Toshiba TCD1304DG). The width of each pixel is 8~$\mu$m, so that the CCD width of $\approx\!30$~mm matches the width of the spatially resolved spectrum emerging from the X-CBG (the height of the CCD chip is 200~$\mu$m). The two CCD chips can be abutted directly to the X-CBG [Fig.~\ref{fig:SystemPhotograph}(d)] and connected to the appropriate electronic circuitry.

To validate the potential for such a dual-band spectrometer, we make use of the supercontinuum source and measure the spectrum directly with a commercial spectrometer (Ocean Optics, S2000). We direct a beam of diameter $\sim\!1$~mm from the source to the input facet of the X-CBG and capture the spatially resolved spectra using the CCD chips on either side of the X-CBG. We plot the measured spectra in each channel as resolved by the X-CBG alongside the reference spectra in Fig.~\ref{fig:SystemPhotograph}(e) and Fig.~\ref{fig:SystemPhotograph}(f). The measured and reference spectra are in good agreement except at the short-wavelength end of the spectrum, which emerges at the far end of the X-CBG as depicted schematically in Fig.~\ref{fig:SystemPhotograph}(a). Note that we have not calibrated the CCD chip to account for the spectral efficiency of the X-CBG.

Further work is needed to optimize the diffraction efficiency of the multiplexed r-CBGs in an X-CBG. In addition to multiplexing two orthogonal r-CBGs in the same volume, another possibility is to multiplex distinct devices along the sample \textit{axis}. This would yield massive multi-functionality in an ultra-compact footprint. For example, two axially multiplexed X-CBGs can provide 4 widely different spectral channels in the same device. Furthermore, the two facets of the device -- the top and bottom facets in Fig.~\ref{fig:SystemPhotograph}(a,b) -- can also be exploited as spectral channels. In this conception, a single device volume can provide massively parallel spectral analysis capabilities. Finally, we note that r-CBGs also provide the opportunity for polarization-sensitive spectral analysis, which we have not exploited here but will report on in more detail in a separate study.

In conclusion, we have realized a novel optical device that we have called an X-CBG, which comprises multiplexed r-CBGs in the same volume. By writing multiple holograms in the same volume, the resulting X-CBG provides the possibility of dual-band spectral analysis in a compact footprint. We realized here several X-CBGs in which we vary the central wavelength and the bandwidth in both spectral channels. Such X-CBGs may pave the way to new applications in miniaturized multi-wavelength and multi-band spectral analysis in fluorescence and nonlinear microscopy, environmental sensing, and portable or handheld devices.

\textbf{Funding:}
U.S. Office of Naval Research (ONR) N00014-17-1-2458 and N00014-20-1-2789.

\textbf{Disclosures:}
The authors declare no conflicts of interest.

\textbf{Data availability}
Data underlying the results presented in this paper are not publicly available at this time but may be obtained from the authors upon reasonable request.

\bibliography{diffraction}

\end{document}